# From the spin eigenmodes of isolated Néel skyrmions to the magnonic bands of a skyrmionic crystal: a micromagnetic study as a function of the strength of both the interfacial Dzyaloshinskii-Moriya and the exchange constants

Mattia Bassotti[1], Raffaele Silvani[2], and Giovanni Carlotti[1,3*]

[1]Dipartimento di Fisica e Geologia, University of Perugia, Via Pascoli, 06123 Perugia, Italy
[2] Istituto Nazionale di Ricerca Metrologica, INRIM, Strada delle Cacce Torino, Italy
[3] Istituto Officina dei Materiali del CNR, c/o Dipartimento di Fisica e Geologia, University of Perugia, Via Pascoli, 06123 Perugia, Italy

*Abstract*— The presence of interfacial Dzyaloshinskii-Moriya interaction (DMI) may lead to the appearance of Néel skyrmions in ferromagnetic films. These topologically protected structures, whose diameter is as small as a few nanometers, can be nowadays stabilized at room temperature and have been proposed for the realization of artificial magnonic crystals and new spintronic devices, such as racetrack memories. In this perspective, it is of utmost importance to analyze their dynamical properties in the GHz range, i.e. in the operation range of current communication devices. Here we exploited the software MuMax3 to calculate the dynamics of Néel skyrmions in the range between 1 and 30 GHz, considering first the eigenmodes of an isolated skyrmion, then the case of two interacting skyrmions and finally a linear chain, representing a one-dimensional magnonic crystal, whose magnonic band structure has been calculated as a function of the strength of both the DMI- and the exchange-constants, namely D and A. The magnonic bands can be interpreted as derived from the eigenmodes of isolated skyrmions, even if hybridization and anti-crossing phenomena occur for specific ranges of values of D and A. Therefore, varying the latter parameters, for instance by a proper choice of the materials and thicknesses, may enable one to fine-tune the permitted and forbidden frequency interval of the corresponding magnonic crystal.

## I. INTRODUCTION

It is well known that magnetic structures characterized by an integer topological charge, such as skyrmions (SKs), have attracted great interest in the last decade for both fundamental reasons and the possible exploitation in information and communication technology (ICT) devices. [1,2]. To this respect, even if SKs were first observed in bulk material lacking inversion symmetry, such as insulating and multiferroic Cu $_2$OSeO$_3$ [3] or β-Mn type Co-Zn-Mn alloy [4], it is very promising the more recent capability of observing Néel SKs in thin films and multilayers, thanks to the presence of chiral magnetic interaction of the Dzyaloshinskii-Moriya form [5,6]. Nowadays, Néel SKs can be stabilized at room temperature, even without the need of an applied external field, so that the path is open to exploit them in real electronic devices. To this aim, intense efforts have been carried on in the last five years to manipulate SKs by external stimuli such as magnetic fields and electrical current and this is an important step towards applications, since it has been envisaged that SKs can be useful for neuromorphic, magnonic, and logic devices [7,8] and that efficient and low power data processing could be achieved exploiting SKs as the elemental carriers of information in racetrack memory devices [9,10]. Here the information flow can be associated with metastable SKs driven along a magnetic strip by pulses of electric current. In this context, it is of utmost importance to analyze also their dynamic properties, because it has been realized that they exhibit a very rich magnetization dynamics in the GHz range of frequencies, reflecting their non-trivial topology. [11,12,13] Such rich and complicated dynamical properties represent interesting perspectives for applications in the fields of spintronics and magnonics, but can also result in unwanted electrical noise in such arrangements [14,15]. In the case of isolated SKs, previous studies showed that there are spin eigenmodes keeping the radial symmetry of the static state (radially symmetric or breathing modes, BR) and eigenmodes with broken radial symmetry (azimuthal modes rotating in the clockwise (CW) and counterclockwise (CCW) directions) [16,17,18]. The radially symmetric modes have no net in-plane magnetization and can be excited only by out-of-plane variable magnetic field. The azimuthal modes, instead, can be excited by an in-plane variable magnetic field. Recently, the experimental proof of the microwave response of skyrmions has been reported [19].

The interest towards the SK dynamics is not only directed towards isolated nanostructures, but also towards arrays of interacting SKs that represent a new class of one- or two-dimensional magnonic crystals (MCs) [20,21,22,23] and where the allowed and forbidden bands can be tuned by external effects, such as external magnetic fields, electric currents, electric fields. In 2015, Ma et al. [20] first reported a numerical study relative to a periodic array of SKs created by nanocontacts carrying a spin current, showing that the propagation of spin waves at specific frequencies could be easily switched on and off by changing the external field and the spin current. In 2016, Mruczkiewicz et al. [21] a theoretical analysis about the dispersion relations of coupled SKs and identified the excited modes in the magnon bands as BR and CW gyrotropic dynamics. More recently, Kim et al. [22,23] investigated the coupled modes in chains of interacting SKs, indicating that these modes can be modulated by the distance between the neighboring elements.

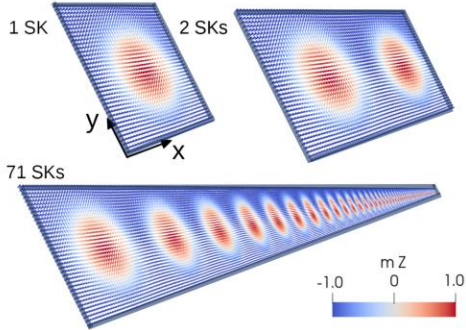

Fig. 1. The three topological systems analyzed in this paper: a single SK, a pair of twin SKs and a linear chain of SKs, representing a one-dimensional MC. The sketches represent region 1 (FeCo) for each of the three system, surrounded by the dark frame that represent region 2 (SmCo).

In this paper we want to analyze the dynamical properties of one-dimensional MCs consisting of linear chain of Nèel SKs, showing how the magnonic bands of collective modes can be described starting from the discrete modes of an isolated SK. Moreover, we want to study how the magnonic bands evolve varying both the DMI constant D and the exchange constant A in a wide range of values. In fact, it has been found that the value of D can be adjusted in a relatively large interval by a proper choice of the heavy metal underlayer and of the ferromagnetic material and thickness. For instance, it may assume values between about 0.1 and 2.5 mJ/m² for the popular Pt/Co system, changing the thicknesses or the overlayer material [24]. On the other hand, also the Heisenberg exchange constant A has been recently shown to assume different values in Pt/Co ultrathin films, ranging from about 5 to 25 pJ/m, also depending on the evaluation methodology [25].

## II. METHODS

Here we exploited the micromagnetic software MuMax3 [26] to calculate the ground state and the dynamics of SKs in the range between 1 and 30 GHz, at zero temperature, in systems with three different levels of complexity, considering first the eigenmodes of an isolated SK, then the case of two interacting SKs and finally a linear chain of 71 units, as sketched in Fig.1. For each simulation, the sample was discretized in cubic cells of side 1 nm and the spin-wave eigenmodes, excited by a pulse of external magnetic field, have been studied as a function of the intensity of the DMI constant D and the exchange constant A.

For the single SK system, the simulated area consists in a single layer of 1 nm thickness, divided in two regions. Region 1 hosts the SK and has dimension of 38x38 nm² with typical micromagnetic parameters of FeCo ($M_s$ = 1.02 MA/m²; $K_u$ = 0.65 kJ/m² [27]).

Region 2 consists in a frame of cells, surrounding region 1, having the typical magnetic parameters of SmCo ($M_s^{(2)}$ = 0.86 MA/m²; $A_{ex}^{(2)}$ = 12 pJ/m; $K_u^{(2)}$ = 17.2 kJ/m² [28]). Such a hard magnetic region is designed to contain the SK and prevents it from touching the edges of the virtual sample and eventually annihilate [29]. The other two systems were simulated repeating the above-described region 1 of 38×38 nm² two times (for the twin of SKs) or 71 times (for the one-dimensional MC) along the x axis (this number is sufficiently high to achieve a proper spatial resolution in the reciprocal space, but sufficiently small to limit the computational effort). As in the single Sk case, the whole simulated region is surrounded by a hard magnetic frame (region 2). For each simulation, SKs were nucleated initializing the magnetization of a circular area in the positive z direction, opposite to the external magnetization pointing along negative z direction, and then relaxing the system for 10 ns. The dynamics was excited applying an external magnetic field pulse: $b(t) = b_0 \sin(2\pi f_0(t-t_0))/2\pi f_0(t-t_0)$ directed perpendicularly to the sample plane, with amplitude $b_0$ = 3 mT, maximum frequency $f_0$ = 50 GHz and, $t_0$ = 1.5 ns, that excites all the frequencies up to a cut off $f_c$ = 50 GHz. The position where the pulse was applied and the direction of the applied field were chosen to efficiently visualize all the eigenmodes in the data analysis. Then, the out of plane component of the reduced magnetization of each cell of region 1 was saved by the software every 10 ps and for a simulated time of 10 ns for the single and twin SK systems, while 25 ns for the SK chain. This corresponds to a frequency resolution of about 0.1 GHz and 0.04 GHz for the two cases, respectively.

The data analysis was made exploiting the fast Fourier transform (fft): frequency spectra were obtained calculating the time-fft of the out-of-plane dynamical magnetization component $m_z$. Snapshots of the magnetization at fixed frequency were obtained setting to zero all the time-fft coefficients corresponding to a frequency different from the chosen one and then re-calculating the magnetization in real space, after inverse-fft. An analogous procedure was used to obtain the spatial profiles at fixed frequency and wave vector for the SK chain analysis (see ref [30] for more details about this methodology).

## III. RESULTS AND DISCUSSION

### A. Eigenmodes of an isolated skyrmion and of a twin of interacting skyrmions

Figs. 2 and 3 present the results relative to the isolated SK and to the twin of coupled SKs placed side to side, respectively.

In the former case, we observed three main modes, whose spatial profiles are shown as colored insets in Fig. 2(a), that have been found also in previous studies of Néel SK dynamics [16, 31].[1] They correspond to the BR mode, whose phase is constant all around the SK and two azimuthal modes with opposite chirality: the gyrotropic counter-clockwise (CCW) mode at low frequency and the CW mode at higher frequency [32]. The latter modes have different frequencies because the CCW mode has the same azimuthal symmetry as the SK topological (gyrotropic) mode and it has large amplitude in the inner domain. On the contrary, the CW mode rotates clockwise, oppositely to the gyrotropic rotation sense and, therefore, it has higher frequency and vanishing amplitude in the inner domain. [33] Animations showing the temporal evolution of the $m_z$ dynamical component for each mode can be found in the supplemental information (SI). The evolution of the mode frequencies with the strength of the DMI constant D and with the exchange constant A are shown in panels (b) and (c) of Fig. 2. It can be seen that they all exhibit a blue frequency-shift on increasing the DMI strength, with a slope that is larger for the modes at higher frequency. Instead, their frequency is almost independent of the exchange constant, showing a slight decrease (increase) for the BR and CCW (CW) mode. On the contrary, it appears also a second-order azimuthal mode, labelled CCW2, whose frequency remarkably decreases (increases) as a function of the DMI constant D (exchange constant A) intensity, crossing the curves relative to the main modes. This mode is characterized by two nodal azimuthal lines, as shown in the profiles seen in the SI videos, so the different behavior of such high-order CCW mode, if compared to CW ones, can be regarded as due to two main causes. First of all, the DMI-induced non reciprocity that is maximum [24] for SWs propagating perpendicularly to the in-plane magnetization, as it happens for modes circulating along the SK circumference. This condition is fulfilled by the azimuthal SWs localized at the SK edge, and nonreciprocity can have an effect on their spectrum. In second place, the CCW2 mode, characterized by two nodal planes at a distance of 10-20 nm, along the circumference of the SK, is in some sense an exchange-dominated mode, because the exchange contribution to the mode frequency is relatively strong if compared to the other contributions. Therefore, it has a frequency rapidly decreasing (increasing) with the DMI constant D (exchange constant A), because the SK radius increases from about 10 to 25 nm when D is lifted from 1.0 to 3.0 mJ/m$^2$ at constant A=9.25 pJ/m (decreases from about 22 to about 12 nm when A is lifted from 8 to 20 pJ/m at constant D=2.0 mJ/m$^2$) so the distance between the azimuthal nodes becomes larger (lower) and the exchange contribution to the mode frequency decreases (increases).

---

[1] The color maps of Figs. 2 and 3 show the phase of the out-of-plane dynamical component $m_z$ and the sum of the moduli of each of the three dynamical components, $m_x$, $m_y$ and $m_z$, obtained by the fft.

If one now moves to the case of two adjacent SKs, we found that each of the main modes discussed for the single SK, namely the BR, CW and CCW modes, undergoes a splitting in two modes, characterized by either a symmetric (S) or antisymmetric (AS) dynamics of the out-of-plane dynamical component $m_z$. This is illustrated in Fig 3(a), where one can see that the amplitude of the splitting is larger for the modes at larger frequency. The physical origin of the splitting can be understood looking at the spatial profiles reported as color insets in Fig. 3(a). The modes behave as coupled oscillators, as already discussed in a previous study [22], where both dipolar interaction and, to a less extent, exchange interaction play a role. For the BR mode one finds at lower frequency the AS-BR mode, characterized by an out-of-phase precession of $m_z$ in the two SKs. On the other hand, for the azimuthal CCW and CW modes the $m_z$ component is out-of-phase in the border region between the two SKs for the S-CCW and S-CW modes and these that lie at lower frequency with respect to the AS counterparts. Moreover, one can notice that the spatial profiles of the modes at lower frequency in each doublet show a sort of attraction in the border region. This is because in this border region the in-plane dynamical components $m_x$ and $m_y$ are much larger than $m_z$, and they are in-phase for the low-frequency modes in each doublet (while they are out-of-phase for the high-frequency modes). As for the dependence of the eigenmode frequencies on either D or A (Fig. 3(b) and (c)), the modes follow a behavior that is similar to the one already observed for the single SK.

## B. Magnonic bands of a linear chains

With the information about the dynamics of both the single SK and the pair of coupled SKs in hands, we can now move towards the analysis of the dynamics of a long chain of 71 interacting SKs, that constitutes an example of an artificial one-dimensional MC with periodicity a=38 nm. As stated in the Method section, we have excited this system by a perpendicular pulsed field applied to the SK placed in the center of the chain and then observed how the dynamics expands all along the chain. Then, from a fft of the temporal evolution of all the discretized cells of the chain, we could obtain the band structure and the spatial profile of the collective eigenmodes. In Fig. 4 we show the obtained band structure for different values of the DMI constant D and of the exchange constant A.

Let us start our analysis from the left panel of Fig. 4 (a), i.e. for relatively small value of the DMI constant D=1 mJ/m$^2$. In this case one can see that there are three well defined low-lying magnonic bands, that can be labelled as the three main modes found in the single and coupled SKs: CCW, BR and CW. This is because we verified that the spatial profiles of these collective modes (not shown here) are very similar, in each element of the chain, to those observed previously for the single SK. The dependence on $k_x$ manifests itself in the fact that at $k_x$=0 these modes are in phase with each-other all over the chain, while for $k_x$=π/a they are out-of-phase in adjacent SKs. For intermediate values of kx, the collective mode has an associated wavelength that involves several SKs, as observed in the calculated spatial profiles corresponding to different values of the wavenumber (not shown here for the lack of space). Remarkably, one may notice that the frequency dispersion is positive for the CCW and the CW modes, while it is negative for the BR mode. This opposite behavior reflects the different frequencies for symmetric and antisymmetric dynamics in adjacent SKs, discussed in the previous section. In the next panels of Fig. 4(a), it is evident that on increasing the value of D the CCW2 band, consistently with what we have seen for the single SK in Fig 2(a), decreases in frequency and crosses the three main magnonic bands, causing hybridization and repulsion phenomena. In particular, when D is lifted from 1 to 1.5 mJ/m$^2$, the CCW2 mode hybridizes with the CW one, while the BR band is lifted and the frequency gap between the BR and the CW bands is reduced from about 1.5 to below 1 GHz. Instead, on further increasing D up to 2.5 mJ/m$^2$, there is a further decrease in frequency of the CCW2 band that exhibit hybridization and anti-crossing with the BR band, so that the spatial profile of the collective mode is not anymore the same within each band in the whole range of $k_x$. Moreover, the amplitude of the forbidden band gap is strongly affected by mode repulsion, so that the gap between the CCW2 and the BR bands significantly increases up to about 2.5 GHz.

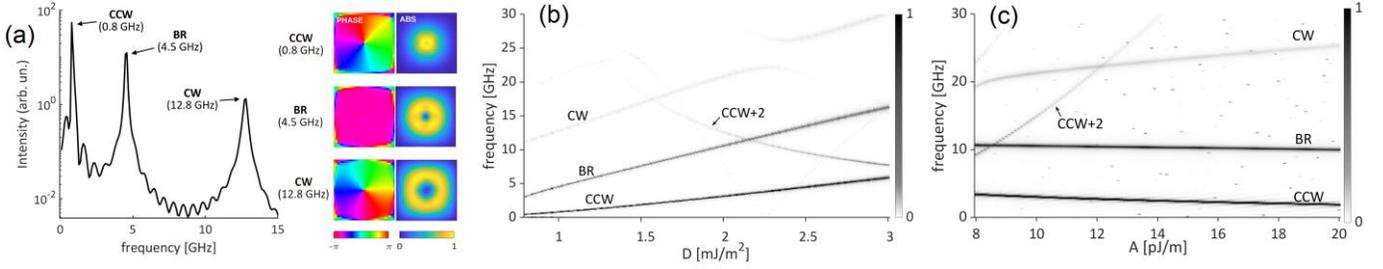

**Fig. 2** (a) Spectrum of the eigenmodes of an isolated SK (A=9.25 pJ/m and D= 1.0 mJ/m$^2$) . The colored insets show the phase of the dynamical $m_z$ component and the sum of the moduli of the dynamical magnetization components. (b) Dependence of the eigenmodes frequencies on the strength of the DMI constant D (in the range 0.8-3.0 mJ/m$^2$, at step of 0.02 mJ/m$^2$) for a fixed value of the exchange constant A=9.25 pJ/m . (c) Dependence of the eigenmodes frequencies on the strength of the exchange constant A (in the range 8-20 pJ/m at step of 0.1 pJ/m) for a fixed value of the DMI constant D=2.0 mJ/m$^2$ .

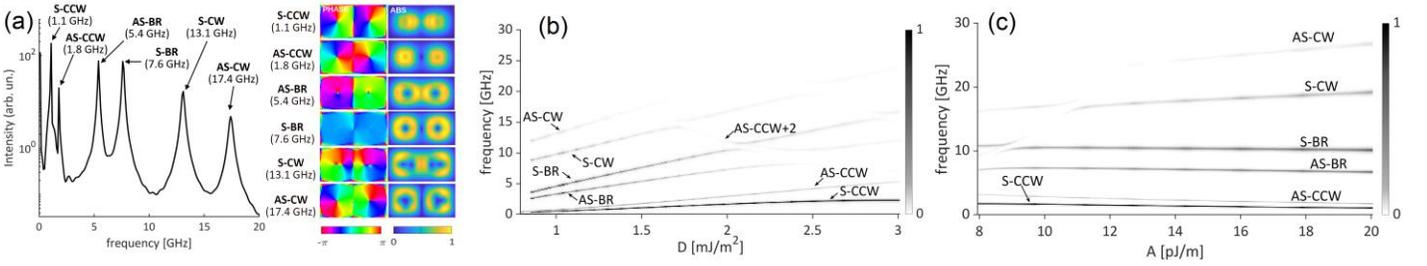

**Fig. 3** (a) Spectrum of the eigenmodes of a twin of interacting SKs (A=9.25 pJ/m and D= 1.5 mJ/m$^2$). The colored insets show the phase of the dynamical $m_z$ component and the sum of the moduli of the dynamical magnetization components. (b) Dependence of the eigenmodes frequencies on the strength of the DMI constant D (in the range 0.8-3.0 mJ/m$^2$, at step of 0.02 mJ/m$^2$) for a fixed value of the exchange constant A=9.25 pJ/m. (c) Dependence of the eigenmodes frequencies on the strength of the exchange constant A (in the range 8-20 pJ/m at step of 0.1 pJ/m) for a fixed value of the DMI constant D=2.0 mJ/m$^2$ .

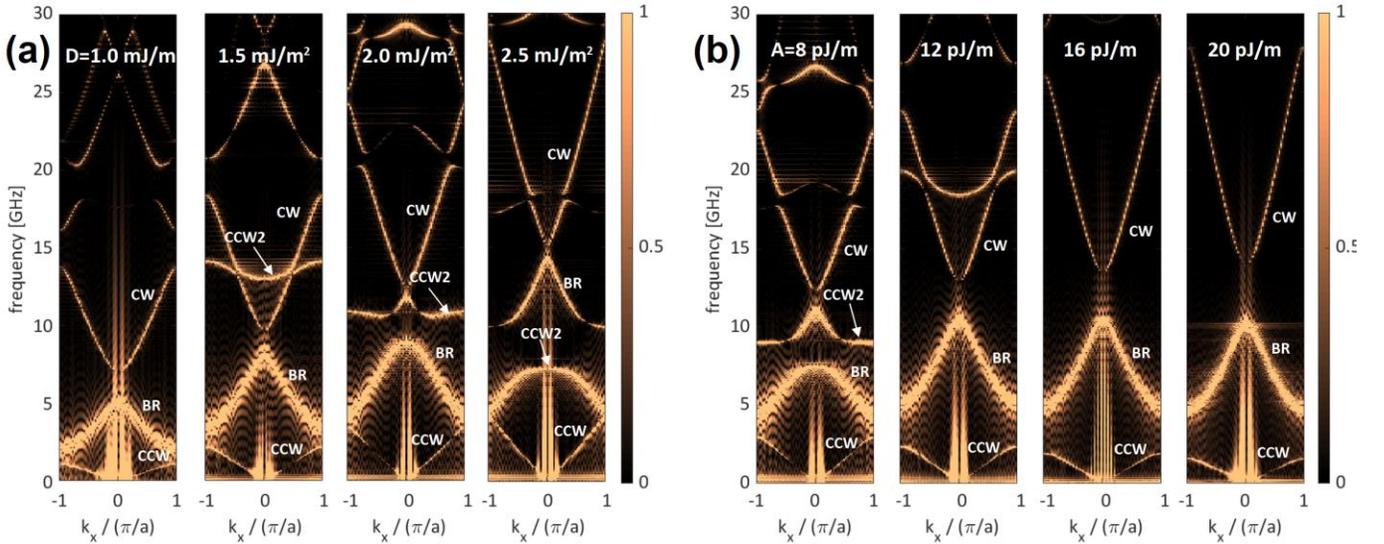

**Fig. 4** (a) Magnonic spectra relative to the linear chains of interacting SKs for different values of the DMI constant D, for a fixed value of the exchange constant A=9.25 pJ/m. (b) Magnonic spectra relative to the linear chains of interacting SKs for different values of the exchange constant A, for a fixed value of the DMI constant D=2 mJ/m$^2$.

Let us now consider the evolution of the magnonic band structure with the strength of the exchange constant A, for a fixed value of D.

In the four panels of Fig. 4(b) it is seen that on increasing the value of A starting from 8 pJ/m causes a simplification of the band structure. This is because the SKs become more rigid and high-order modes that are more sensitive to the exchange, such as the CCW2, are pushed to very high frequencies, so that a relatively simple band structure is recovered for A larger than about 15 pJ/m. For such relatively large values of A, the forbidden gaps between the BR and both the CCW and CW bands are lifted above 2-3 GHz and one comes back to the three well separated bands CCW, BR and CW.

## IV. CONCLUSION

We have shown that the band structure of one-dimensional magnonic crystals consisting of interacting Néel SKs can be interpreted in terms of magnonic bands that derive from the main eigenmodes of isolated SK, separated by forbidden band gaps. However, this simple picture is a good approximation only for either relatively small values of the DMI constant D or for relatively large values of the exchange constant A. In general, instead, the magnonic band structure is characterized by hybrid collective excitations, with anticrossing and band repulsion phenomena, caused by the dipolar interaction among the elemental constituents of the chain. Consequently, the amplitude of the permitted frequency bands and of the forbidden gaps are remarkably sensitive to variations of both D and A in the range of values typical of current ferromagnetic materials. This suggests that both parameters could be in principle exploited for fine-tuning the permitted and forbidden operation frequencies of future devices based on such kind of skyrmionics magnonic crystals.

## ACKNOWLEDGMENT

This work has been carried on within the project 17FUN08 TOPS that has received funding from the EMPIR programme co-financed by the Participating States and from the European Union's Horizon 2020 research and innovation programme.